\documentclass[aps,prl,twocolumn,superscriptaddress,showpacs,preprintnumbers,amsmath,amssymb]{revtex4}
\usepackage{graphicx}
\usepackage{dcolumn}
\graphicspath{{ps}}

\begin{document}
\preprint{\vbox{ \hbox{   }
    \hbox{BELLE Preprint 2009-19}
    \hbox{KEK Preprint 2009-19}
}}

\title{ \quad\\[1.0cm] Measurement of $CP$ asymmetries in $B^0 \to K^0\pi^0$ decays}
\affiliation{Budker Institute of Nuclear Physics, Novosibirsk}
\affiliation{Chiba University, Chiba}
\affiliation{University of Cincinnati, Cincinnati, Ohio 45221}
\affiliation{The Graduate University for Advanced Studies, Hayama}
\affiliation{Gyeongsang National University, Chinju}
\affiliation{Hanyang University, Seoul}
\affiliation{University of Hawaii, Honolulu, Hawaii 96822}
\affiliation{High Energy Accelerator Research Organization (KEK), Tsukuba}
\affiliation{Hiroshima Institute of Technology, Hiroshima}
\affiliation{Institute of High Energy Physics, Chinese Academy of Sciences, Beijing}
\affiliation{Institute of High Energy Physics, Vienna}
\affiliation{Institute of High Energy Physics, Protvino}
\affiliation{INFN - Sezione di Torino, Torino}
\affiliation{Institute for Theoretical and Experimental Physics, Moscow}
\affiliation{J. Stefan Institute, Ljubljana}
\affiliation{Kanagawa University, Yokohama}
\affiliation{Institut f\"ur Experimentelle Kernphysik, Universit\"at Karlsruhe, Karlsruhe}
\affiliation{Korea University, Seoul}
\affiliation{Kyoto University, Kyoto}
\affiliation{Kyungpook National University, Taegu}
\affiliation{\'Ecole Polytechnique F\'ed\'erale de Lausanne (EPFL), Lausanne}
\affiliation{Faculty of Mathematics and Physics, University of Ljubljana, Ljubljana}
\affiliation{University of Maribor, Maribor}
\affiliation{Max-Planck-Institut f\"ur Physik, M\"unchen}
\affiliation{University of Melbourne, School of Physics, Victoria 3010}
\affiliation{Nagoya University, Nagoya}
\affiliation{Nara Women's University, Nara}
\affiliation{National Central University, Chung-li}
\affiliation{National United University, Miao Li}
\affiliation{Department of Physics, National Taiwan University, Taipei}
\affiliation{H. Niewodniczanski Institute of Nuclear Physics, Krakow}
\affiliation{Nippon Dental University, Niigata}
\affiliation{Niigata University, Niigata}
\affiliation{University of Nova Gorica, Nova Gorica}
\affiliation{Novosibirsk State University, Novosibirsk}
\affiliation{Osaka City University, Osaka}
\affiliation{Panjab University, Chandigarh}
\affiliation{University of Science and Technology of China, Hefei}
\affiliation{Seoul National University, Seoul}
\affiliation{Shinshu University, Nagano}
\affiliation{Sungkyunkwan University, Suwon}
\affiliation{University of Sydney, Sydney, New South Wales}
\affiliation{Tata Institute of Fundamental Research, Mumbai}
\affiliation{Excellence Cluster Universe, Technische Universit\"at M\"unchen, Garching}
\affiliation{Tohoku Gakuin University, Tagajo}
\affiliation{Tohoku University, Sendai}
\affiliation{Department of Physics, University of Tokyo, Tokyo}
\affiliation{Tokyo Institute of Technology, Tokyo}
\affiliation{Tokyo Metropolitan University, Tokyo}
\affiliation{Tokyo University of Agriculture and Technology, Tokyo}
\affiliation{IPNAS, Virginia Polytechnic Institute and State University, Blacksburg, Virginia 24061}
\affiliation{Yonsei University, Seoul}
  \author{M.~Fujikawa}\affiliation{Nara Women's University, Nara} 
  \author{Y.~Yusa}\affiliation{IPNAS, Virginia Polytechnic Institute and State University, Blacksburg, Virginia 24061} 
  \author{J.~Dalseno}\affiliation{Max-Planck-Institut f\"ur Physik, M\"unchen}\affiliation{Excellence Cluster Universe, Technische Universit\"at M\"unchen, Garching} 
  \author{M.~Hazumi}\affiliation{High Energy Accelerator Research Organization (KEK), Tsukuba} 
  \author{K.~Sumisawa}\affiliation{High Energy Accelerator Research Organization (KEK), Tsukuba} 
  \author{H.~Aihara}\affiliation{Department of Physics, University of Tokyo, Tokyo} 
  \author{K.~Arinstein}\affiliation{Budker Institute of Nuclear Physics, Novosibirsk}\affiliation{Novosibirsk State University, Novosibirsk} 
  \author{V.~Aulchenko}\affiliation{Budker Institute of Nuclear Physics, Novosibirsk}\affiliation{Novosibirsk State University, Novosibirsk} 
  \author{T.~Aushev}\affiliation{\'Ecole Polytechnique F\'ed\'erale de Lausanne (EPFL), Lausanne}\affiliation{Institute for Theoretical and Experimental Physics, Moscow} 
  \author{T.~Aziz}\affiliation{Tata Institute of Fundamental Research, Mumbai} 
% ...

\collaboration{The Belle Collaboration}

\begin{abstract}
We report measurements of $CP$ violation parameters in $B^0 \to K^0 \pi^0$ decays based on a data sample of $657 \times 10^6 B\bar{B}$ pairs collected with the Belle detector at the KEKB $e^+ e^-$ asymmetric-energy collider. We use $B^0 \to K^0_S \pi^0$ decays for both mixing-induced and direct $CP$ violating asymmetry measurements and $B^0 \to K^0_L \pi^0$ decays for the direct $CP$ violation measurement. The $CP$ violation parameters obtained are $\sin 2 \phi_1^{\rm eff} = +0.67 \pm 0.31\mbox{(stat)} \pm 0.08 \mbox{(syst)}$ and $\mathcal{A}_{K^0 \pi^0} = +0.14 \pm 0.13\mbox{(stat)} \pm 0.06 \mbox{(syst)}$. 
The branching fraction of $B^0 \to K^0 \pi^0$ decay is measured to be $\mathcal{B}(B^0 \to K^0 \pi^0) = (8.7\pm0.5 (\rm{stat.})\pm0.6 (\rm{syst.}) )\times 10^{-6}$. The observed $\mathcal{A}_{K^0 \pi^0}$ value differs by 1.9 standard deviations from the value expected from an isospin sum rule.
\end{abstract}


\maketitle

\tighten

{\renewcommand{\thefootnote}{\fnsymbol{footnote}}}
\setcounter{footnote}{0}
Decays of $B$ mesons mediated by $b \to s$ penguin amplitudes play an important role in both measuring the Standard Model (SM) parameters and in probing new physics. In the SM, $CP$ violation arises from a single irreducible Kobayashi-Maskawa (KM) phase~\cite{KMphase}, in the weak-interaction quark-mixing matrix. In the decay sequences $\Upsilon(4S) \to B^0\bar{B}^0 \to f_{CP}f_{\rm tag}$, where one of the $B$ mesons decays at time $t_{CP}$ to a $CP$ eigenstate $f_{CP}$ and the other decays at time $t_{\rm tag}$ to a final state $f_{\rm tag}$ that distinguishes between $B^0$ and $\bar{B}^0$, the decay rate has a time dependence given by 
\begin{eqnarray}
\displaystyle \mathcal{P}(\Delta t) &=& \frac{e^{-|\Delta t|/\tau_{B^0}}}{4\tau_{B^0}} \bigg[ 1+q\cdot\big[\mathcal{S}_f\sin(\Delta m_d \Delta t) \nonumber\\ 
 & & ~~~~~~~~~~~~~~~~~~~~~+ \mathcal{A}_f\cos(\Delta m_d \Delta t)\big] \bigg]. 
\label{PDFsig}
\end{eqnarray}
Here, $\mathcal{S}_f$ and $\mathcal{A}_f$ are parameters that describe mixing-induced and direct $CP$ violation, respectively, $\tau_{B^0}$ is the $B^0$ lifetime, $\Delta m_d$ is the mass difference between the two $B^0$ mass eigenstates, $\Delta t = t_{CP} - t_{\rm tag}$, and the $b$-flavor charge, $q = +1 (-1)$ when the tagged $B$ meson is a $B^0 (\bar{B}^0)$. 
The SM predicts $\mathcal{S}_f = -\xi_f \sin 2 \phi_1$ and $\mathcal{A}_f \simeq 0$ to a good approximation for most of the decays that proceed via $b \to s q\bar{q}~(q = c, s, d, u)$ quark transitions \cite{sqq_penguin}, where $\xi_f = +1(-1)$ corresponds to $CP$-even (-odd) final states and $\phi_1$ is an angle of the unitarity triangle. The final state $K^0_S \pi^0$ is a $CP$ eigenstate with $CP$ eigenvalue $\xi_f = -1$ while $K^0_L \pi^0$ is a $CP$ eigenstate with $\xi_f = +1$. 

However, even within the SM, in $B^0 \to K^0 \pi^0$ decay modes, both $\mathcal{S}_{K^0\pi^0}$ and $\mathcal{A}_{K^0\pi^0}$ could be shifted due to the contribution of a color-suppressed tree diagram that has a $V_{ub}$ coupling~\cite{shift_Vub}. The resulting effective parameter $\sin 2 \phi_1^{\rm eff}$ can be evaluated in the $1/m_b$ expansion and/or using SU(3) flavor symmetry~\cite{shift_Sterm}, whereas the shift in $\mathcal{A}_{K^0\pi^0}$ is predicted by applying an isospin sum rule to the recent measurements of $B$ meson decays into $K\pi$ final states~\cite{shift_Aterm}. The sum rule for the decay rates gives the following relation to within a few percent precision determined by SU(2) flavor symmetry ~\cite{gronau_sum_rule},
\begin{eqnarray}
\mathcal{A}_{K^+\pi^-} + \mathcal{A}_{K^0\pi^+}\frac{\mathcal{B}(K^0\pi^+)\tau_{B^0}}{\mathcal{B}(K^+\pi^-)\tau_{B^+}}= \nonumber \\
\mathcal{A}_{K^+\pi^0}\frac{2\mathcal{B}(K^+\pi^0)\tau_{B^0}}{\mathcal{B}(K^+\pi^-)\tau_{B^+}} +\mathcal{A}_{K^0\pi^0}\frac{2\mathcal{B}(K^0\pi^0)}{\mathcal{B}(K^+\pi^-)}.
\label{Eq_sumrule}
\end{eqnarray}
Here, $\mathcal{B}$ represents the branching fraction of a decay mode. Since the branching fractions and $CP$ asymmetries of other $B \to K \pi$ decay modes have been measured with good precision~\cite{belle_Kpi, babar_Kpi}, $\mathcal{A}_{K^0\pi^0}$ is constrained in this framework with a small error. Therefore, a significant discrepancy between the measured and expected values of $\mathcal{A}_{K^0 \pi^0}$ would indicate a new physics contribution to the sum rule. The expected uncertainty in $\mathcal{A}_{K^0 \pi^0}$ can be reduced by improved measurement of the $B^0 \to K^0 \pi^0$ branching fraction. Furthermore, recent measurements that show an unexpectedly large difference between $\mathcal{A}_{K^+\pi^-}$ and $\mathcal{A}_{K^+\pi^0}$~\cite{babar_Kpi, belle_nature} makes an improved measurement of $\mathcal{A}_{K^0\pi^0}$ particularly interesting. In this paper, in addition to the $B^0 \to K^0_S \pi^0$ mode, we measure the $CP$ asymmetry in $B^0 \to K^0_L \pi^0$ decay for the first time, in order to maximize sensitivity to the direct $CP$ violation parameter, $\mathcal{A}_{K^0\pi^0}$.

At the KEKB asymmetric-energy $e^+e^-$ (3.5~GeV on 8~GeV) collider~\cite{KEKB}, the $\Upsilon(4S)$ is produced with a Lorentz boost of $\beta\gamma = 0.425$ nearly along the direction opposite to the positron beam line ($z$-axis). Since $B^0$ and $\bar{B}^0$ mesons are approximately at rest in the $\Upsilon(4S)$ center-of-mass system (CM), $\Delta t$ can be determined from the displacement in $z$ between the $f_{CP}$ and $f_{\rm tag}$ decay vertices: $\Delta t \simeq (z_{CP}-z_{\rm tag})/(\beta\gamma c) \equiv \Delta z/(\beta\gamma c)$. For $K_S^0\pi^0$ decays, the vertex position of the $CP$ side is determined from the $K^0_S$ decay products and the $K^0_S$ mesons are required to decay within the silicon vertex detector (SVD) for the time dependent $CP$ violation measurement. Since we cannot obtain vertex information from $K_L^0\pi^0$ decays, only the parameter $\mathcal{A}_{K^0\pi^0}$ is measured by comparing the decay rates of $B^0 \to K^0_L\pi^0$ and $\bar{B}^0 \to K^0_L\pi^0$. The subset of $B^0 \to K^0_S\pi^0$ events for which we cannot obtain $\Delta t$ from the decay vertex reconstruction are treated similarly.

Previous measurements of $CP$ violation in $B^0 \to K^0_S \pi^0$ decay have been reported by Belle~\cite{kspi0_belle06} and BABAR~\cite{kspi0_babar09}. The previous result from Belle was based on $532 \times 10^6$ $B\bar{B}$ pairs. In this report, all results are based on a data sample that contains $657 \times 10^6$ $B\bar{B}$ pairs, collected  with the Belle detector at the KEKB operating at the $\Upsilon(4S)$ resonance. 

The Belle detector is a large-solid-angle magnetic spectrometer that consists of SVD, a 50-layer central drift chamber (CDC), an array of aerogel threshold Cherenkov counters (ACC), a barrel-like arrangement of time-of-flight scintillation counters (TOF), and an electromagnetic calorimeter (ECL) comprised of CsI(Tl) crystals located inside a superconducting solenoid coil that provides a 1.5~T magnetic field.  An iron flux-return located outside of the coil is instrumented to detect $K_L^0$ mesons and to identify muons (KLM).  The detector is described in detail elsewhere~\cite{Belle}. Two configurations of the inner detectors were used. A 2.0 cm -radius beampipe and a 3-layer silicon vertex detector were used for the first sample of $152 \times 10^6 B\bar{B}$ pairs, while a 1.5 cm -radius beampipe, a 4-layer silicon detector and a small-cell inner drift chamber were used to record the remaining $505 \times 10^6 B\bar{B}$ pairs~\cite{svd2}.  

Charged particle tracks are reconstructed with the SVD and CDC. Photons are identified as isolated ECL clusters that are not matched to any charged particle track. We reconstruct $\pi^0$ candidates from pairs of photons that have energies larger than the following thresholds: 50 MeV for the barrel region and 100 MeV for the endcap regions, where the barrel region covers the polar angle range $32^\circ < \theta < 130^\circ$, and the endcap regions cover the forward and backward regions. The invariant mass of reconstructed $\pi^0$'s are required to be in the range between 0.115 GeV$/c^2$ and 0.152 GeV$/c^2$. We reconstruct $K^0_S$ candidates from pairs of oppositely charged tracks having invariant mass between 0.480 GeV/$c^2$ and 0.516 GeV/$c^2$, which corresponds to three standard deviations in a Gaussian fit to the signal Monte Carlo (MC) samples. The flight direction of each $K^0_S$ candidate is required to be consistent with the direction of its vertex displacement with respect to the interaction point (IP). Candidate $K^0_L$ mesons are selected from ECL and/or KLM hit patterns that are not associated with any charged tracks and consistent with the presence of a shower induced by a $K^0_L$ meson \cite{KLselection}.

For reconstructed $B^0 \to K^0_S \pi^0$ candidates, we identify $B$ meson decays using the beam-constrained mass $M_{\rm bc} \equiv \sqrt{(E^{\rm CM}_{\rm beam})^2 - (p^{\rm CM}_B)^2}$ and the energy difference $\Delta E \equiv E^{\rm CM}_B - E^{\rm CM}_{\rm beam}$, where $E^{\rm CM}_{\rm beam}$ is the beam energy in the CM, and $E^{\rm CM}_B$ and $p^{\rm CM}_B$ are the CM energy and momentum of the reconstructed $B$ candidate, respectively. The signal candidates used for measurements of $CP$ violation parameters are selected by requiring 5.27 GeV/$c^2 < M_{\rm bc} <$ 5.29 GeV/$c^2$ and $-0.15$ GeV $< \Delta E <$ 0.1 GeV. For $B^0 \to K^0_L \pi^0$ candidates, we can only measure the flight direction of the $K^0_L$, so $M_{\rm bc}$ is calculated by assuming the parent $B^0$ to be at rest in the CM. The signal is selected by requiring $M_{\rm bc} >$ 5.255 GeV/$c^2$. In the $B^0 \to K^0_S \pi^0$ analysis, if there are multiple candidates, we select the candidate that has the smallest $\chi^2$ of the $\pi^0$ mass-constrained fit for the daughter photons. In the $B^0 \to K^0_L \pi^0$ case, the candidate having the smallest $\cos\theta_{\rm exp}$ is chosen, where $\theta_{\rm exp}$ is the angle between the measured $K^0_L$ flight direction and that expected from the $\pi^0$ momentum assuming the parent $B^0$ to be at rest in the CM frame.

The dominant background for the signal is from continuum $e^+e^- \to u\bar{u}, d\bar{d}, s\bar{s}$ or $c\bar{c}$ events. To distinguish the spherical $B\bar{B}$ signal events from these jet-like backgrounds, we combine a set of variables that characterize the event topology, i.e., modified-Fox-Wolfram moments~\cite{SFW1, SFW2, KSFW}, into a signal (background) likelihood variable $\mathcal{L}_{\rm sig (bkg)}$, and impose requirements on the likelihood ratio $\mathcal{R}_{\rm s/b} \equiv \mathcal{L}_{\rm sig}/(\mathcal{L}_{\rm sig}+\mathcal{L}_{\rm bkg})$: $\mathcal{R}_{\rm s/b} > 0.3$ and $\mathcal{R}_{\rm s/b}>0.5$ for $B^0 \to K^0_S \pi^0$ and $B^0 \to K^0_L \pi^0$ candidates, respectively. 

The $b$-flavor of the accompanying $B$ meson is determined from inclusive properties of particles that are not associated with the reconstructed $B^0 \to K^0 \pi^0$ decays. To represent the tagging information, we use two parameters, the $b$-flavor charge, $q$ and its quality factor, $r$~\cite{TaggingNIM}. The parameter $r$ is an event-by-event, MC determined flavor-tagging dilution factor that ranges from $r = 0$ for no flavor discrimination to $r = 1$ for unambiguous flavor assignment. For events with $r > 0.1$, the wrong tag fractions for six $r$ intervals, $w_l$ $(l = 1$-$6)$, and their differences between $B^0$ and $\bar{B}^0$ decays, $\Delta w_l$, are determined using a high-statistics control sample of semi-leptonic and hadronic $b \to c$ decays~\cite{b2s2005, b2c2004}. If $r < 0.1$, we set the wrong tag fraction to 0.5, in which case the accompanying $B$ meson does not provide tagging information and such events are not used for the $CP$ violation parameter measurement. The total effective tagging efficiency is estimated to be $0.29\pm0.01$, where ``effective'' means a summation over the products of tagging efficiency and $r^2$ of all types of tags used. 

The vertex position for the $B^0 \to K^0_S \pi^0$ decay is reconstructed using charged pions from the $K^0_S$ decay and an IP constraint~\cite{vertexres}. Each charged pion track is required to have more than 1(2) hit(s) on SVD $r-\phi~(z)$ strips. The $f_{\rm tag}$ vertex is obtained with well-reconstructed tracks that are not assigned to the $B^0 \to K^0_S \pi^0$ decay. 

\begin{figure}[htb]
  \includegraphics[height=120pt,width=120pt]{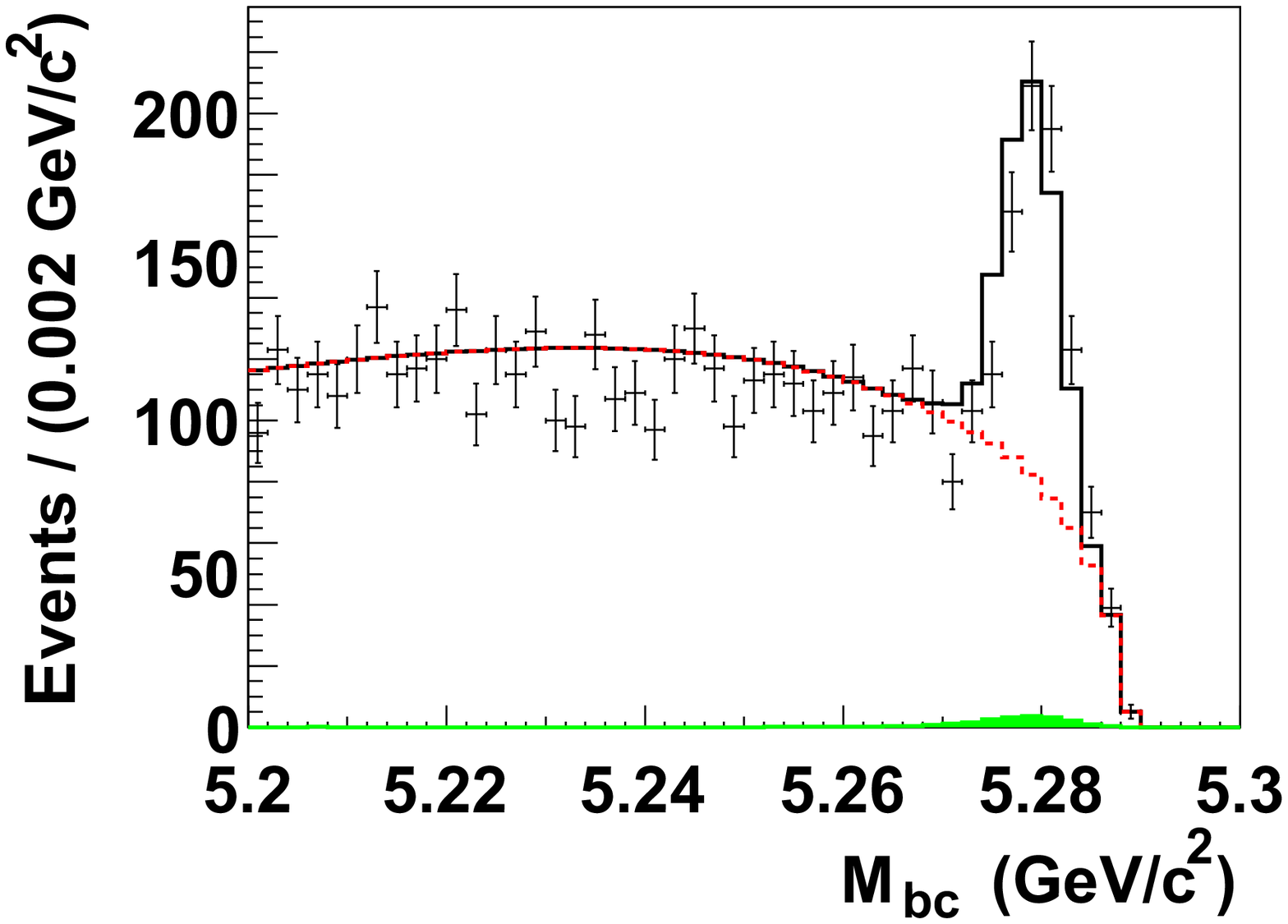}
  \includegraphics[height=120pt,width=120pt]{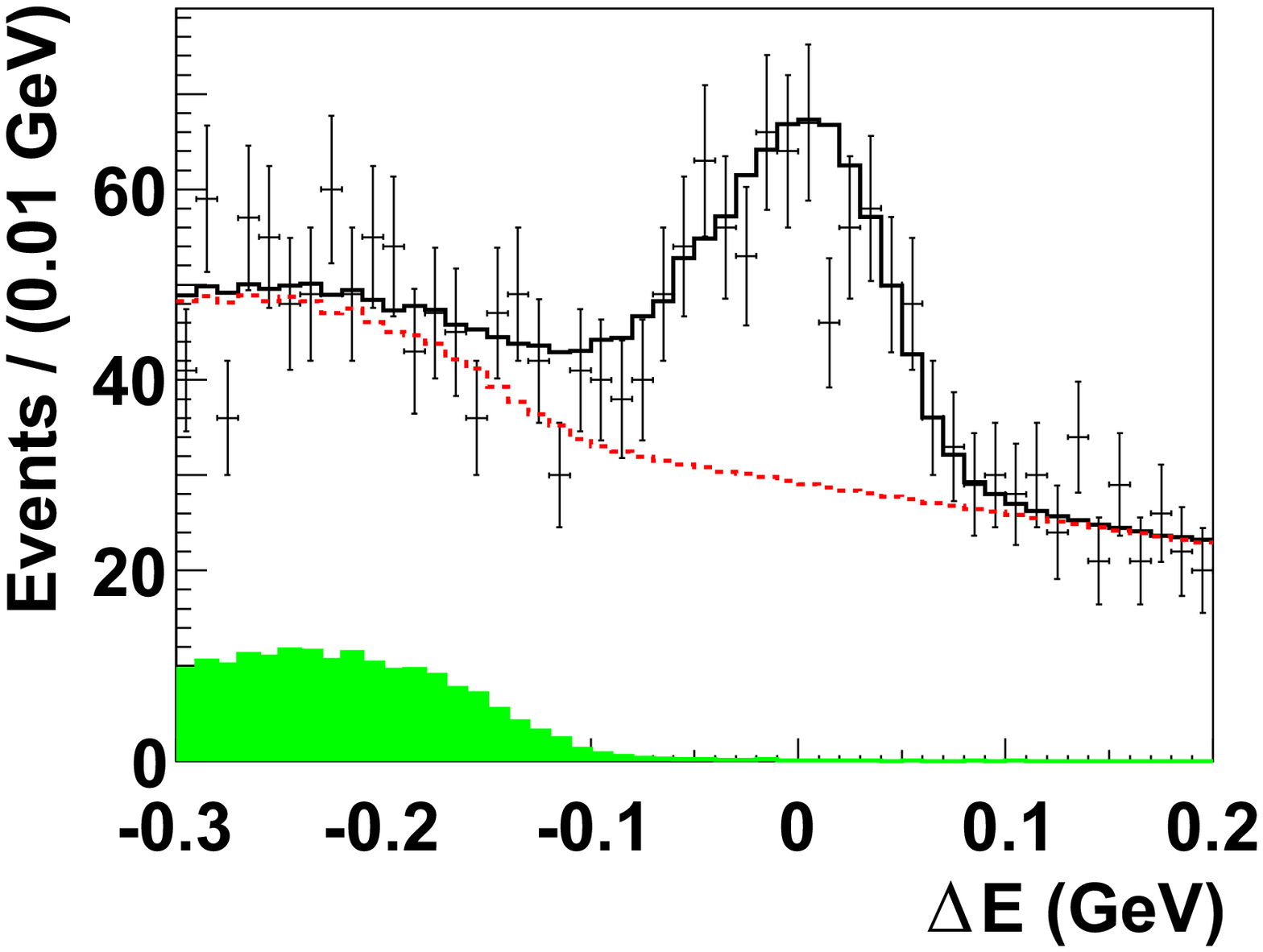}\\
  \includegraphics[height=120pt,width=120pt]{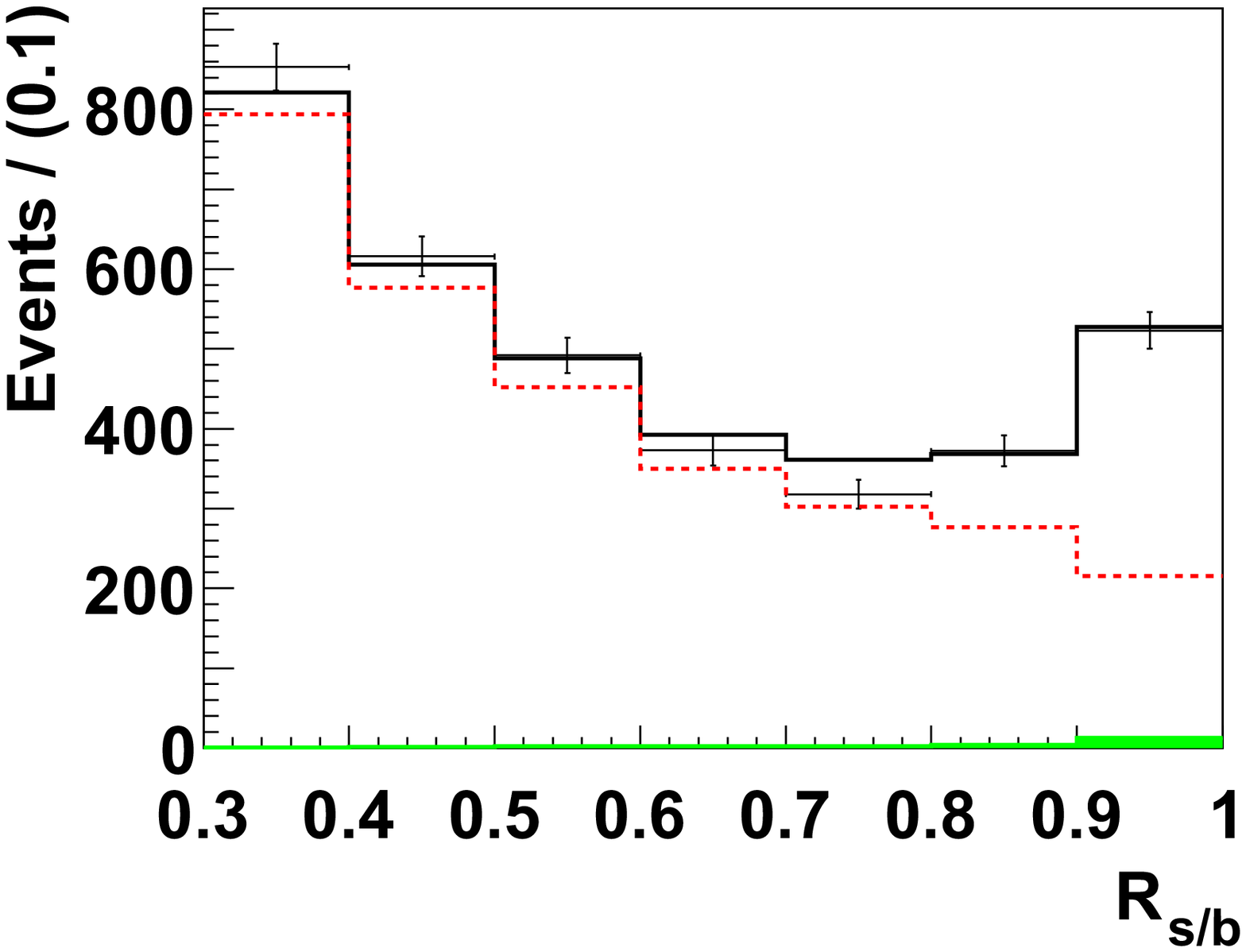}
\caption{$M_{\rm bc}$-$\Delta E$-$\mathcal{R}_{\rm s/b}$ fit projections of $B^0 \to K^0_S \pi^0$ candidates. The open histogram with the solid curve shows the fit result, the filled histogram is the $B\bar{B}$ background, and the dashed histogram is the sum of continuum and $B\bar{B}$ backgrounds. Each plot requires signal enhanced conditions for the other variables: 5.27 GeV/$c^2 < M_{\rm bc} <$ 5.29 GeV/$c^2$, $-0.15$ GeV $< \Delta E <$ 0.1 GeV and $\mathcal{R}_{\rm s/b}>0.7$.}
  \label{fig_kspi0_yield}
\end{figure}

\begin{figure}[htb]
  \includegraphics[height=120pt,width=120pt]{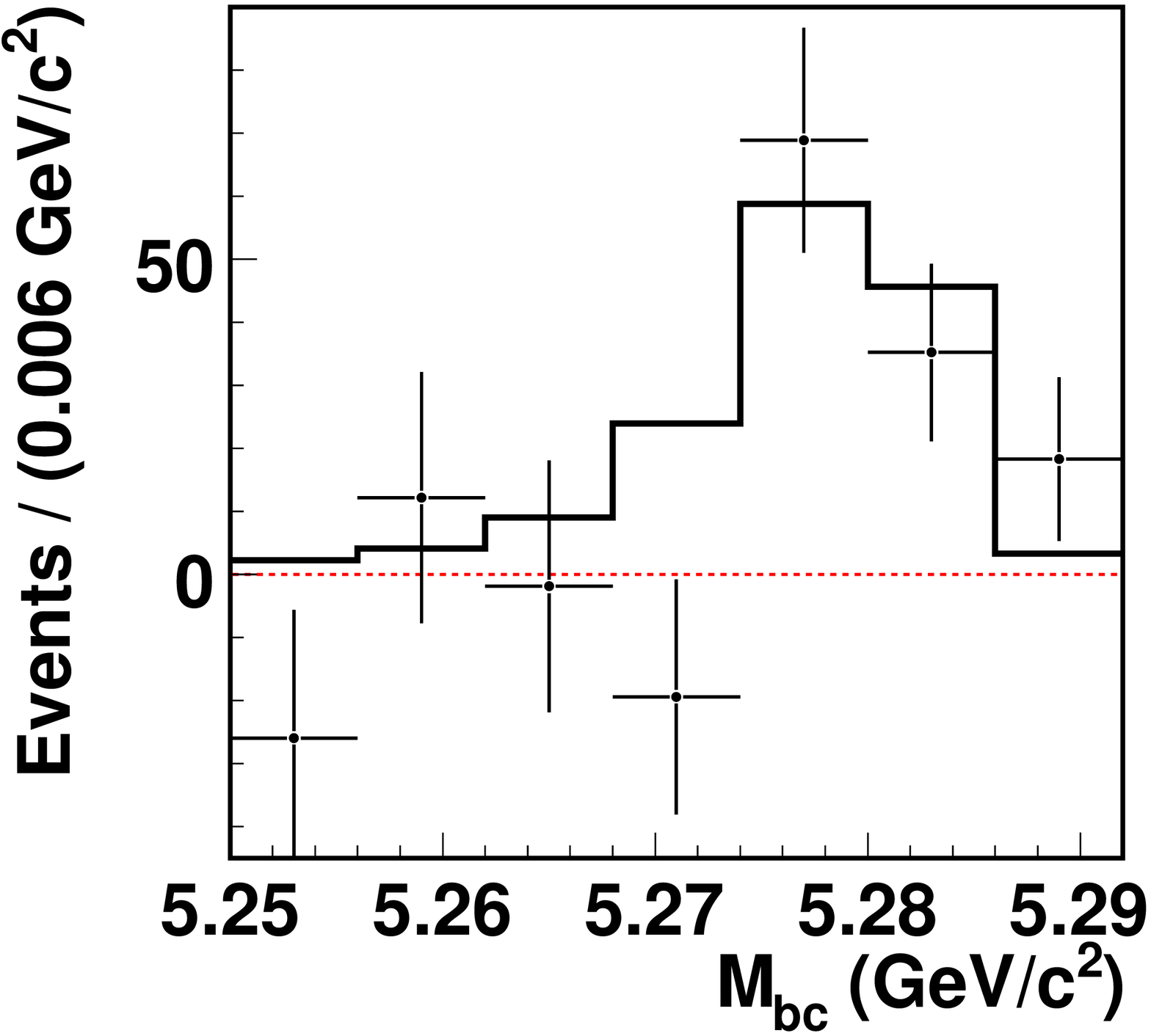}
  \includegraphics[height=120pt,width=120pt]{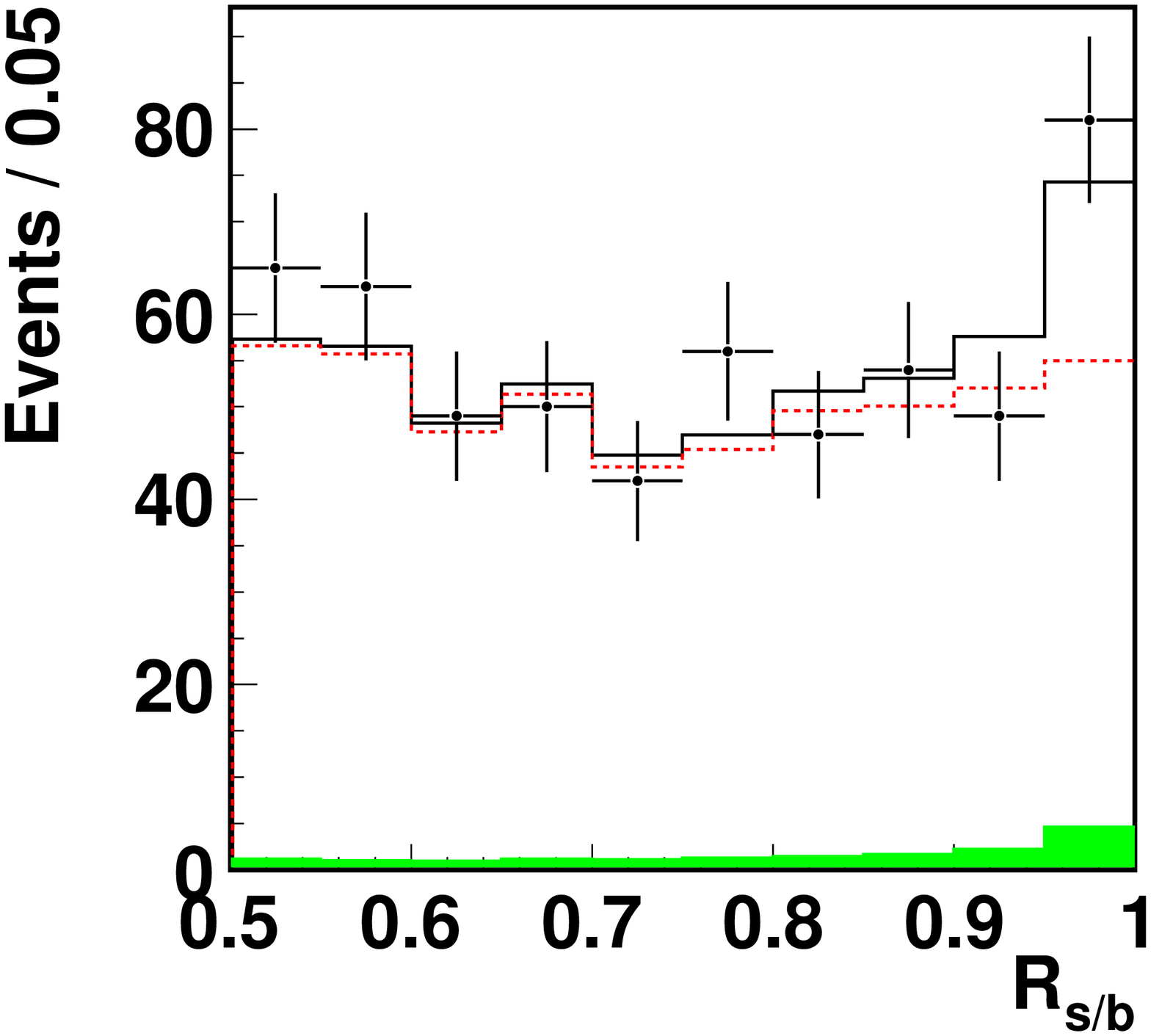}
\caption{$B^0 \to K^0_L \pi^0$ fit projections for events with good tags: background subtracted $M_{\rm bc}$ (left) and $\mathcal{R}_{\rm s/b}$ (right). The open solid histogram shows the fit result. The filled histogram is the $B\bar{B}$ background and the open dashed histogram is the sum of continuum and $B\bar{B}$ background.}
  \label{fig_klpi0_yield}
\end{figure}

Figures~\ref{fig_kspi0_yield} and \ref{fig_klpi0_yield} show the distribution of the selection variables for $B^0 \to K^0_S \pi^0$ and $B^0 \to K^0_L \pi^0$ candidates. The signal yields are obtained from multi-dimensional extended unbinned maximum likelihood fits to these distributions. The $M_{\rm bc}$, $\Delta E$ and $\mathcal{R}_{\rm s/b}$ signal shapes for $B^0 \to K^0_S \pi^0$ are modeled with three-dimensional histograms determined from MC, while the continuum background shapes in $M_{\rm bc}$ and $\Delta E$ are modeled with an ARGUS function~\cite{ARGUS func} and a linear function, respectively, whose shape and normalization are floated in the fit. The data from a sideband region (5.20 GeV/$c^2 < M_{\rm bc} <$ 5.26 GeV/$c^2$ and 0.05 $< \Delta E <$ 0.20 GeV) are used to determine the continuum background shape in $\mathcal{R}_{\rm s/b}$. For $B^0 \to K^0_L \pi^0$, the signal shape in $M_{\rm bc}$ is determined from MC samples and the continuum background shape is modeled with an ARGUS function. The signal shape in $\mathcal{R}_{\rm s/b}$ is determined from MC simulation. The continuum $\mathcal{R}_{\rm s/b}$ shape is determined using $\Upsilon (4S)$ off-resonance data. The shape of each variable in the $B\bar{B}$ background component is modeled using MC events. The signal yield is extracted in each $r$-bin for $B^0 \to K^0_L \pi^0$ candidates with $r$-dependent $\mathcal{R}_{\rm s/b}$ shapes. For $B^0 \to K^0_L \pi^0$, the ratio of $B\bar{B}$ background to signal is evaluated from MC simulated events and the $B\bar{B}$ background contribution is then fixed according to that of the signal in the fit. 

We perform a fit to $B^0 \to K^0_S \pi^0$ candidates using a signal shape with correction factors (to account for small differences between data and MC) obtained from $B^+ \to K^+ \pi^0$. The signal yield is $634 \pm 34$, where the error is statistical only. The average signal detection efficiency is calculated from MC to be ($22.3 \pm 0.1$)\%. We obtain a $B^0 \to K^0 \pi^0$ branching fraction of $(8.7\pm0.5\pm0.6)\times 10^{-6}$ using only $B^0 \to K^0_S \pi^0$ candidates, where the first error is statistical and the second is systematic. The systematic uncertainty for the $B^0 \to K^0 \pi^0$ branching fraction is estimated by varying the correction factors obtained from $B^+ \to K^+ \pi^0$ by $\pm 1\sigma$ ($+3.6$/$-2.4$\%) and varying histogram probability density functions (PDF's) bin-by-bin by $\pm 2\sigma$ ($+1.5$/$-1.6$\%). Uncertainties in the number of $B\bar{B}$ pairs (1.4\%), MC statistics (0.2\%), $K^0_S$ (4.9\%) and $\pi^0$ reconstruction efficiencies (2.8\%) are also included.

The signal yield of $B^0 \to K^0_L \pi^0$ is $285\pm52$, where the error is statistical only. We evaluate the systematic uncertainty for the $B^0 \to K^0_L \pi^0$ signal yield by smearing the PDF shapes of $M_{\rm bc}$, $\mathcal{R}_{\rm s/b}$ and $r$ used in the fit. The dominant contribution is from the continuum background shape and the total systematic error is 20\%. Taking into account both statistical and systematic errors, the significance of $B^0 \to K^0_L \pi^0$ is 3.7$\sigma$.


We determine $\sin 2 \phi_1^{\rm eff}$ and $\mathcal{A}_{K^0_S\pi^0}$ for $B^0 \to K^0_S \pi^0$ by performing an unbinned maximum-likelihood fit to the observed $\Delta t$ distribution. The PDF expected for the signal distribution, $\mathcal{P}(\Delta t; \sin 2 \phi_1^{\rm eff}, \mathcal{A}_{K^0_S\pi^0}, q, w_l, \Delta w_l)$, is given by Eq. $($\ref{PDFsig}$)$ fixing $\tau_{B^0}$ and $\Delta m_d$ to their world averages~\cite{PDG} and incorporating the effect of wrong flavor assignment. The distribution is convolved with the proper-time interval resolution function, $R_{\rm sig}(\Delta t)$, which takes into account the finite vertex resolution. The resolution is determined by a multi-parameter fit to the $\Delta t$ distributions of high-statistics control samples of $B^0 \to J/\psi K^0_S$ decays~\cite{b2s2005, b2c2004}, where the $K^0_S$ is used for vertex reconstruction. We determine the following likelihood for each event, 
\begin{eqnarray}
\displaystyle P_i & = & (1-f_{\rm ol}) \int \bigg[ f_{\rm sig} \mathcal{P}_{\rm sig}(\Delta t')R_{\rm sig} (\Delta t_i - \Delta t') \nonumber\\
 & + & (1 - f_{\rm sig})\mathcal{P}_{\rm bkg}(\Delta t')R_{\rm bkg}(\Delta t_i - \Delta t')\bigg]d(\Delta t')\nonumber\\
 & + & f_{\rm ol} P_{\rm ol}(\Delta t_i).
\label{UML_LR}
\end{eqnarray}
The signal probability, $f_{\rm sig}$, depends on $r$ and is calculated in each region on an event-by-event basis as a function of $M_{\rm bc}, \mathcal{R}_{\rm s/b}$ and, where applicable, $\Delta E$ from the shapes given in  Figs.~\ref{fig_kspi0_yield} and \ref{fig_klpi0_yield}. $\mathcal{P}_{\rm bkg}$ is a PDF for continuum and $B\bar{B}$ backgrounds. The background PDF's are determined from $M_{\rm bc}$ and $\Delta E$ sideband data for continuum and, MC and data for $B\bar{B}$. The term, $P_{\rm ol}(\Delta t)$ is a broad Gaussian function that represents a small outlier component with a fraction $f_{\rm ol}$~\cite{b2s2005, b2c2004}. The free parameters in the final fits are $\sin 2 \phi_1^{\rm eff}$ and $\mathcal{A}_{K^0_S\pi^0}$, which are determined by maximizing the likelihood function $L = \Pi_i P_i(\Delta t_i; \sin 2 \phi_1^{\rm eff}, \mathcal{A}_{K^0_S\pi^0})$ where the product is over all events.

The $B^0 \to K^0_L \pi^0$ and $B^0 \to K^0_S \pi^0$ candidates that do not have vertex information are only used for the determination of $\mathcal{A}_{K^0\pi^0}$. Since $\Delta t$ vanishes by integration, Eq. (\ref{UML_LR}) becomes simpler:
\begin{eqnarray}
P_i = f_{\rm sig}\mathcal{P}_{\rm sig}(q) + (1-f_{\rm sig})\mathcal{P}_{\rm bkg}(q),
\end{eqnarray}
where $\mathcal{P}_{\rm bkg}(q = \pm 1) = 0.5$ since we assign no tag information for the continuum background meaning that the number of events tagged as $q = +1$ and $q = -1$ are equal. Since no $CP$ violation is expected in the background outlier component, we include the $f_{\rm ol}$ term in the $\mathcal{P}_{\rm bkg}$ PDF. The signal PDF is obtained by integrating the time-dependent decay rate Eq. (\ref{PDFsig}) from $-\infty$ to $+\infty$:
\begin{eqnarray}
\mathcal{P}_{\rm sig}(q; \mathcal{A}_{K^0_L\pi^0}) & = & \frac{1}{2}\bigg[ 1 + \frac{q \mathcal{A}_{K^0\pi^0}}{1+\tau_{B^0}^2\Delta m_d^2}\bigg].
\end{eqnarray}

\begin{figure}[htb]
\includegraphics[scale=0.37]{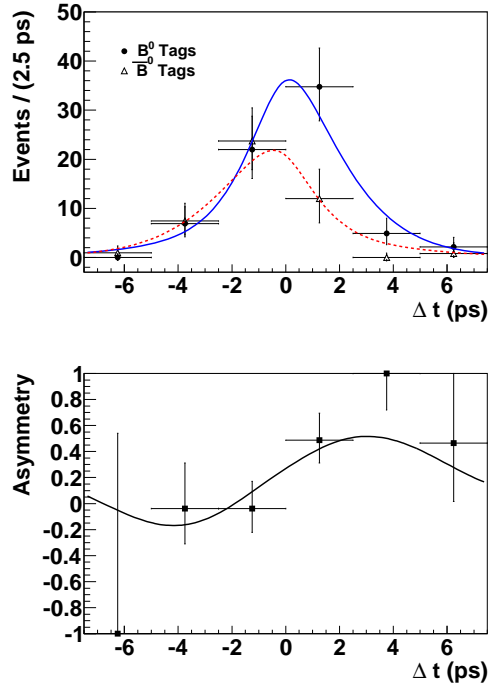}
\caption{The top plot shows the background subtracted $\Delta t$ distribution for $B^0$ and $\bar{B^0}$ tags where the solid (broken) curve represents the $\Delta t$ curve for $B^0$ ($\bar{B^0}$) in the good tag region $0.5 < r \leq 1.0$. The bottom plot shows the background subtracted asymmetry defined as $(N^{\rm Sig}_{\bar{B}^0} - N^{\rm Sig}_{{B^0}})/(N^{\rm Sig}_{\bar{B}^0} + N^{\rm Sig}_{{B^0}})$ in each $\Delta t$ bin where $N^{\rm Sig}_{{B}^0}$ ($N^{\rm Sig}_{\bar{B^0}})$) is the $B^0$ ($\bar{B^0}$) signal yield extracted in that $\Delta t$ bin. The solid curve shows the $CP$ asymmetry result expected from the fit.}
  \label{fig_KSassym}
\end{figure}

We obtain the fit results $\sin 2 \phi_1^{\rm eff} = +0.67\pm0.31$ and $\mathcal{A}_{K^0\pi^0} = +0.14\pm0.13$ for $B^0 \to K^0 \pi^0$. Fits to individual modes yield $\sin 2 \phi_1^{\rm eff} = +0.67\pm0.31$ and $\mathcal{A}_{K^0_S\pi^0} = +0.15\pm0.13$ for $B^0 \to K^0_S \pi^0$, and $\mathcal{A}_{K^0_L\pi^0} = -0.01\pm0.45$ for $B^0 \to K^0_L \pi^0$, where the errors are statistical only. Fig.~\ref{fig_KSassym} shows the background subtracted $\Delta t$ distributions for $B^0$ and $\bar{B}^0$ tags as well as the asymmetry for $B^0 \to K^0_S \pi^0$ candidates.
The dominant sources of systematic errors are summarized in Table \ref{tab_syst}. The systematic uncertainty from wrong tag fractions, physics parameters, resolution function, background $\Delta t$ and background fractions are studied by varying each parameter by its error. A possible fit bias is examined by fitting a large number of pseudo-experiments. The systematic uncertainty for the vertex reconstruction is estimated by changing the charged track selection criteria. The dominant effect for $\Delta \mathcal{A}_{K^0 \pi^0}$ comes from misalignment between the SVD and CDC. The tag side interference is evaluated from pseudo-experiments in which the effect of possible $CP$ violation in $B^0 \to f_{\rm tag}$ decays is taken into account~\cite{TSI}.
\begin{table}[htb]
\caption{ Systematic uncertainties in $\sin 2 \phi_1^{\rm eff}$ and $\mathcal{A}_{K^0\pi^0}$.}
\label{tab_syst}
\begin{tabular}
 {@{\hspace{0.5cm}}l@{\hspace{0.5cm}}  @{\hspace{0.5cm}}c@{\hspace{0.5cm}} @{\hspace{0.5cm}}c@{\hspace{0.5cm}}}
\hline \hline
Source &  $\Delta \sin 2 \phi_1^{\rm eff}$ & $\Delta \mathcal{A}_{K^0\pi^0}$\\
\hline
Wrong tag fraction          & 0.007 & 0.005\\
Physics parameters          & 0.007 & 0.001\\
Resolution function         & 0.063 & 0.007\\
Background $\Delta t$ shape & 0.015 & 0.006\\
Background fraction         & 0.029 & 0.022\\
Possible fit bias           & 0.010 & 0.020\\
Vertex reconstruction       & 0.013 & 0.022\\
Tag side interference       & 0.014 & 0.054\\
\hline
Total                       & 0.077 & 0.064\\
\hline \hline
\end{tabular}
\end{table}
As a cross-check, we fit the $B^0$ lifetime using the same event sample that is used for the $B^0\to K^0_S \pi^0$ $CP$ violation parameter measurement and obtain $\tau_{B^0} = 1.46 \pm 0.18$ ps, which is consistent with the PDG world average~\cite{PDG}.

In summary, we use $B^0 \to K^0_S \pi^0$ decays to measure the branching fraction and $CP$ violation parameters for $B^0 \to K^0 \pi^0$. We use $B^0 \to K^0_L \pi^0$ decays to measure the direct $CP$ violation parameter. Our results are
\begin{eqnarray}
\mathcal{B}(B^0 \to K^0 \pi^0) &=& (8.7 \pm 0.5 \pm 0.6) \times 10^{-6}\\
\mathcal{A}_{K^0 \pi^0} &=& +0.14 \pm 0.13 \pm 0.06 \\
\sin 2 \phi_1^{\rm eff} &=& +0.67 \pm 0.31 \pm 0.08,
\end{eqnarray}
where the first and second errors listed are statistical and systematic, respectively. These results are consistent with previous measurements~\cite{kspi0_belle06, kspi0_babar09}; the value for the branching fraction is the most precise single measurement to-date. We test the isospin sum rule (Eq. \ref{Eq_sumrule}) by inserting our measured values for the branching fraction and $\mathcal{A}_{K^0\pi^0}$. For the other parameters we use the most recent world average values \cite{HFAG}. We find the isospin relationship to be only marginally satisfied; the level of disagreement is 1.9$\sigma$. Specifically, the difference between our measurement of $\mathcal{A}_{K^0\pi^0} \times \mathcal{B}(K^0\pi^0)$ and that predicted by Eq. \ref{Eq_sumrule} is 1.9$\sigma$.

We thank the KEKB group for the excellent operation of the
accelerator, the KEK cryogenics group for the efficient
operation of the solenoid, and the KEK computer group and
the National Institute of Informatics for valuable computing
and SINET3 network support.  We acknowledge support from
the Ministry of Education, Culture, Sports, Science, and
Technology (MEXT) of Japan, the Japan Society for the 
Promotion of Science (JSPS), and the Tau-Lepton Physics 
Research Center of Nagoya University; 
the Australian Research Council and the Australian 
Department of Industry, Innovation, Science and Research;
the National Natural Science Foundation of China under
contract No.~10575109, 10775142, 10875115 and 10825524; 
the Department of Science and Technology of India; 
the BK21 and WCU program of the Ministry Education Science and
Technology, the CHEP SRC program and Basic Research program (grant No.
R01-2008-000-10477-0) of the Korea Science and Engineering Foundation,
Korea Research Foundation (KRF-2008-313-C00177),
and the Korea Institute of Science and Technology Information;
the Polish Ministry of Science and Higher Education;
the Ministry of Education and Science of the Russian
Federation and the Russian Federal Agency for Atomic Energy;
the Slovenian Research Agency;  the Swiss
National Science Foundation; the National Science Council
and the Ministry of Education of Taiwan; and the U.S.\
Department of Energy.
This work is supported by a Grant-in-Aid from MEXT for 
Science Research in a Priority Area ("New Development of 
Flavor Physics"), and from JSPS for Creative Scientific 
Research ("Evolution of Tau-lepton Physics").

\end{document}